\documentclass[review,number,sort&compress]{elsart}

\usepackage{ifpdf}
\usepackage{graphicx,amssymb}
\usepackage{amssymb}
\ifpdf
\usepackage[%
  pdftitle={SPIDER Identification},%
  pdfauthor={Diego Gruyer},%
  pdfstartview=FitH,%
  bookmarks=true,%
  bookmarksopen=true,%
  breaklinks=true,%
  colorlinks=true,%
  linkcolor=blue,anchorcolor=blue,%
  citecolor=blue,filecolor=blue,%
  menucolor=blue,pagecolor=blue,%
  urlcolor=blue]{hyperref}
\else
\usepackage[%
  breaklinks=true,%
  colorlinks=true,%
  linkcolor=blue,anchorcolor=blue,%
  citecolor=blue,filecolor=blue,%
  menucolor=blue,pagecolor=blue,%
  urlcolor=blue]{hyperref}
\fi

\makeatletter
\def\elsartstyle{%
    \def\normalsize{\@setfontsize\normalsize\@xiipt{14.5}}
    \def\small{\@setfontsize\small\@xipt{13.6}}
    \let\footnotesize=\small
    \def\large{\@setfontsize\large\@xivpt{18}}
    \def\Large{\@setfontsize\Large\@xviipt{22}}
    \skip\@mpfootins = 18\p@ \@plus 2\p@
    \normalsize
}
\@ifundefined{square}{}{\let\Box\square}
\makeatother

\usepackage{epsfig}
\usepackage{graphicx}
\usepackage{color}
\usepackage{textcomp}
\usepackage{tikz}
\usepackage{amsmath}
\usepackage{amssymb}
\usepackage{graphicx}
\usepackage{subfigure}
\usepackage{enumerate}
\usepackage{multirow}

\newcommand{\Dee}{$\Delta E - E$ }

\newcommand{\Mev}{\,MeV/A }

\begin{document}

%\linenumbers

\begin{frontmatter}
\title{Semi-automatic charge and mass identification in two-dimensional matrices}

\author[fir,gan]{D. Gruyer\corauthref{cor}}
\ead{diego.gruyer@fi.infn.it}
\author[gan]{E. Bonnet}
\author[gan]{A. Chbihi}
\author[gan]{J.D. Frankland}
\author[fir,uni]{S. Barlini}
\author[ipno]{B. Borderie}
\author[lpc]{R. Bougault}
\author[esp]{J. A. Due\~nas}
\author[pol]{A. Kordyasz}
\author[pol2]{T. Kozik}
\author[lpc]{N. Le Neindre}
\author[lpc]{O. Lopez}
\author[lpc,rom]{M. P\^arlog}
\author[fir,uni]{G. Pastore}
\author[fir]{S. Piantelli}
\author[fir,uni]{S. Valdr\'e}
\author[ipno,cat]{G. Verde}
\author[lpc]{E. Vient}

\address[fir]{INFN -- Sezione di Firenze, via Sansone 1, I-50019 Sesto Fiorentino, Italy}
\address[gan]{Grand Acc\'el\'erateur National d'Ions Lourds (GANIL), CEA/DRF--CNRS/IN2P3, Bvd. Henri Becquerel, 14076 Caen, France}
\address[uni]{Dipartimento di Fisica, Universit\'a di Firenze, I-50019 Sesto Fiorentino, Italy}
\address[ipno]{Institut de Physique Nucl\'eaire, CNRS/IN2P3, Univ. Paris-Sud, Universit\'e Paris-Saclay, Orsay, France}
\address[lpc]{Normandie Univ, ENSICAEN, UNICAEN, CNRS/IN2P3, LPC Caen, 14000 Caen, France}
\address[esp]{Departamento de Ingenier\'ia El\'ectrica, Escuela T\'ecnica Superiorde Ingenier\'ia, Universidad de Huelva, 21819 Huelva, Spain}
\address[pol]{Heavy Ion Laboratory, University of Warsaw, ul. Pasteura 5A, 02-093 Warsaw, Poland.}
\address[pol2]{Faculty of Physics, Astronomy and Applied Computer Science, Jagiellonian University, 30-348 Cracow, Poland}
\address[rom]{National Institute for Physics and Nuclear Engeneering, RO-76900 Bucharest-M\`agurele, Romania}
\address[cat]{INFN -- Sezione di Catania, 64 Via Santa Sofia, I-95123 Catania, Italy}

\corauth[cor]{Corresponding author.}

\begin{abstract}
  This article presents a new semi-automatic method for charge and mass identification in two-dimensional matrices.
  The proposed algorithm is based on the matrix's properties and uses as little information as possible on the global
  form of the identification lines, making it applicable to a large variety of matrices, including
  Particular attention has been paid to the implementation in a suitable graphical environment, so that only two mouse-clicks
  are required from the user to calculate all initialization parameters.
  Example applications to recent data from both INDRA and FAZIA telescopes are presented.
\end{abstract}

\end{frontmatter}

\section{Introduction}

In the intermediate energy regime, violent heavy-ion collisions produce many nuclear species
with a large range of charge ($Z$), mass ($A$) and kinetic energy ($E_k$) \cite{WCI06,Borderie2008Nuclear}. 
Studying this kind of reactions requires detectors with almost $4\pi$ solid angle coverage,
high granularity, low energy thresholds, large dynamic range in energy and capable of
characterizing reaction products on an event by event basis. 
The first generation of $4\pi$ multi-detectors focused
on complete collection of charged particles produced in a reaction \cite{STRACENER1990485,Pouthas1995418,KWIATKOWSKI1995571,SARANTITES1996418,PAGANO2004504}, 
providing little isotopic information for heavy fragments ($Z>5$).
More recently detectors have evolved to provide isotopic resolution for a broader range of products \cite{Wuenschel2009578,epjaFazia},
by improving existing detectors and identification techniques, or developing new methods such as the Pulse Shape Analysis (PSA) in silicon detectors \cite{Barlini2009644,Bardelli2011272}.  

Such multi-detectors are generally made of telescopes, stacks of detector material layers, 
measuring the energy lost by charged particles in the different stages.
Several combinations of detectors have been used for this purpose, such as ionization
chambers (IC), silicon detectors (Si), plastic scintillators, and thallium-activated cesium-iodide scintillators
(CsI(Tl)). When a charged particle passes through such a telescope, its charge, mass, and kinetic energy
determine the number of detectors it can cross before stopping, and the energy loss in the different layers. 
Charged particles are then identified by plotting the energy loss in one or several layers of the telescope ($\Delta E$)
versus the residual energy released in the detector in which the particle is stopped ($E$). 
Within this representation, called \Dee matrix, different particles populate identification lines 
characteristic of their charge and mass (see for example Fig. \ref{fig:spid}(a)). 

Two main methods are then used to identify such particles:
\begin{enumerate}[(i)]
 \item Interactive drawing of lines in order to discriminate between the ridges corresponding to a given charge and/or mass. 
       Particles are then identified from their relative distance between pairs of ridge lines.
 \item Fit of a limited set of ridge lines with a functional describing the relation between 
       $\Delta E$ and $E$, in which $Z$ and $A$ enter as parameters 
       \cite{Goulding1964New,Butler1970Identification,TassanGot2002New}. In this case, particle identification is obtained 
       by inversion of the functional for given $\Delta E$ and $E$, in order to extract $Z$ and possibly $A$.
\end{enumerate}

The first method is probably more powerful and allows to face any situation, but it suffers two main limitations:
it does not provide any extrapolation in regions of low statistics, and it is time consuming because each line has to be accurately drawn. 
The second method suffers less from this inconvenience, since only a subset of ridge lines have to be drawn by hand, but may still become problematic when using multidetectors composed of thousands of identification telescopes.
In addition, existing functionals are generally not accurate enough to reproduce isotopic lines over a large range of elements.

With increasing numbers of identification matrices to treat which include information
on increasing numbers of individual isotopes of different elements,
it becomes essential to develop automatic or semi-automatic methods to extract identification lines in \Dee matrices.
The need for automation was already evident with the advent of the first large charged particle
arrays, and some methods were developed at that time \cite{Benkirane1995Contextual,MorhacIdentification}. However, these methods have never been used for
large scale identification grid production, mainly due to lack of computer resources. 
Other specific examples can be found in the literature \cite{Morelli2010Automatic,Dudouet2013Comparison}.

The evolution of computer resources, and the availability of powerful libraries dedicated to large scale data analysis
\cite{Brun1997ROOT,kaliveda} allow us to consider new types of algorithms.
In this article we present a new method, called SPIDER identification 
(for Spider Particle Identification in \Dee Representation)
for semi-automatic ridge line determination in two-dimensional matrices.
This method has been developed avoiding as much as possible the use of \textit{a priori} information on the exact form of identification lines,
in order to be applicable to a large variety of identification matrices.
Particular attention has been payed to the implementation in a suitable graphical environment.
The extracted lines can then be directly used to identify charged particles (i), 
or set as an input of a functional fit (ii).

\section{SPIDER Identification}

\subsection{Algorithm}

\begin{figure*}[h]
\begin{center}
\includegraphics[width=.49\linewidth]{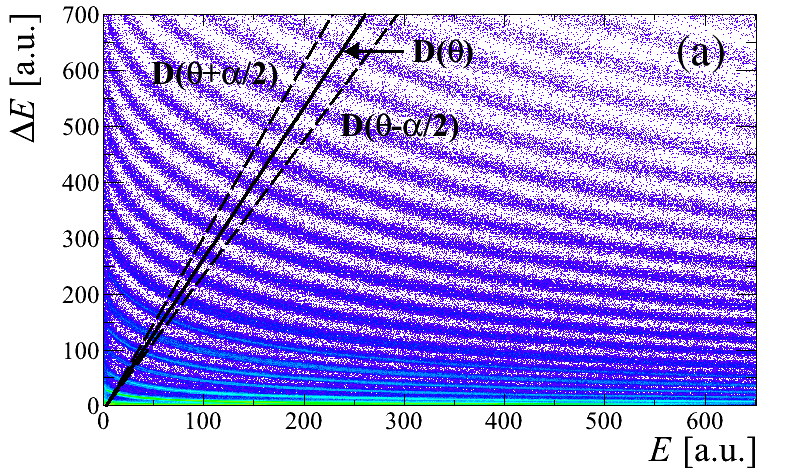}
\includegraphics[width=.49\linewidth]{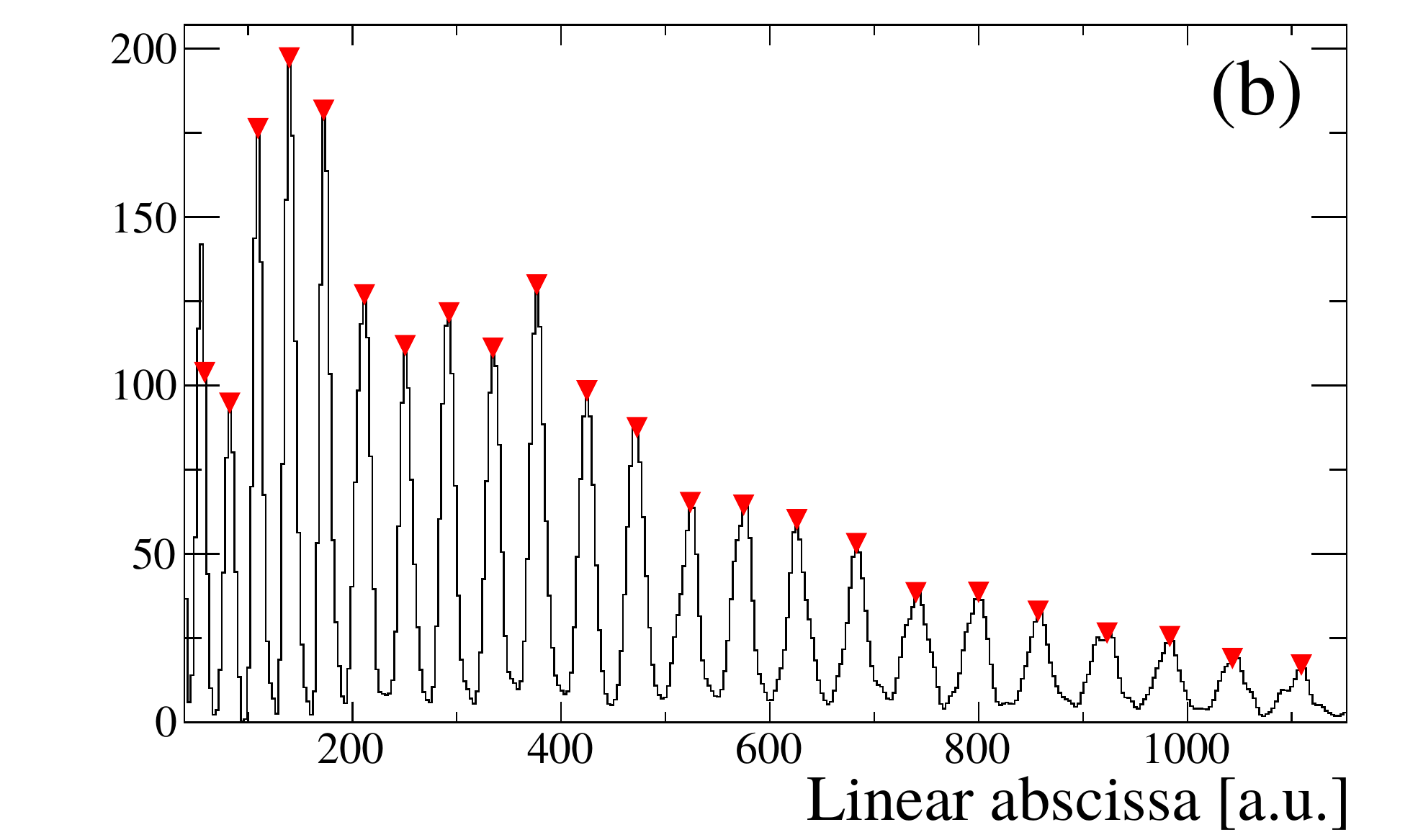}\\
\includegraphics[width=.49\linewidth]{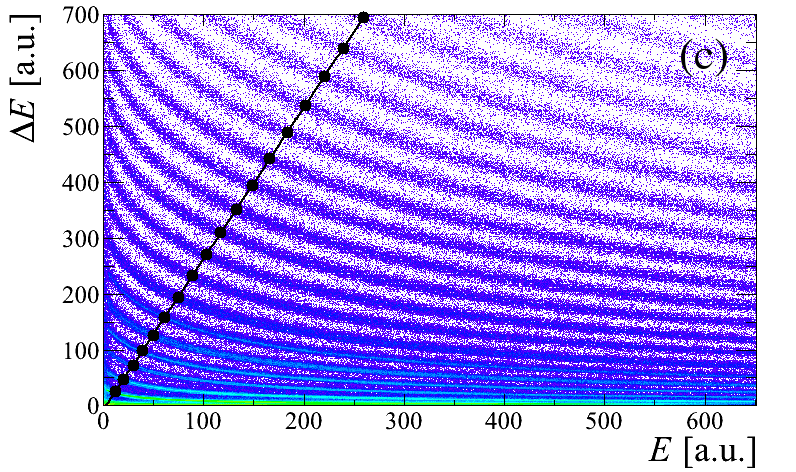}
\includegraphics[width=.49\linewidth]{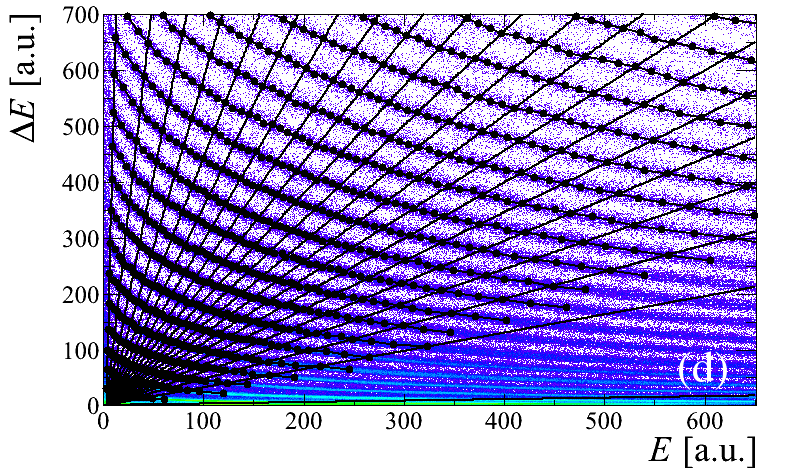}
\caption{
Illustration of the different steps of the SPIDER Identification:
(a) definition of the projection; 
(b) projection on line $D(\theta)$ and localization of maxima;
(c) positioning of the maxima on the \Dee matrix;
(d) weaving of the spiderweb.}
\label{fig:spid}
\end{center}
\end{figure*}

Determining ridge lines in two-dimensional matrices ($x$, $y$) is a hard task, whereas powerful
algorithms for peak localization in $N$-dimensional matrices are available \cite{Morhac1997Efficient}.
The main idea of the present method is then to transform our problem into a problem
of peak localization in one-dimensional histograms.
To do so, we have to project a part of the matrix onto a relevant axis. 
It is the shape of the identification lines and their relative population that guided the choice of this projection.
The one-dimensional histogram shown in Fig. \ref{fig:spid}(b) is obtained by projecting
all points between  $D(\theta-\alpha/2)$ and $D(\theta+\alpha/2)$ onto the straight line $D(\theta)$, passing through $(x_0,y_0)$ and making an angle $\theta$ with respect
to $Ox$ (see Fig. \ref{fig:spid}(a) where $x$ stands for $E$ and $y$ for $\Delta E$). 
Each peak appearing on this projection corresponds to the intersection between $D(\theta)$ and a ridge line of a given $Z$ and $A$\footnote{The ridge lines of individual isotopes are indistinguishable in these data, due to insufficient resolution of the $\Delta E$ detectors. In this case ions of different $Z$ populate broad ridges around the mean value $<A>$ of their isotopic distribution.}. 
The angle of the first projection, $\theta_0$, and the pedestal coordinates $(x_0,y_0)$
are input parameters of the algorithm. 

To make peaks appear on Fig. \ref{fig:spid}(b), the binning of the histogram should be chosen carefully. 
The number of bins $n_b$ of the projection is defined as:
\begin{equation}
n_b = d_{\theta} \times \rho (\theta) \times \beta,
\end{equation}
with
  $d_{\theta}$ the length of the projection, %$D(\thetMorhac1997Efficienta)$ segment included in the matrix range,
  $\beta$ a binning parameter to be provided by the user, and % : %à partir de la position du pivot, du piédestal et de $Z_0$ :
  $\rho(\theta)$ an internal parameter given by:
\begin{equation}
\rho(\theta) =
\left\{
\begin{array}{ll}
    \sqrt{2} &\text{ if $\theta>\theta_0$}\\
    1 &\text{ if $\theta=\theta_0$}\\
    5/4 &\text{ if $\theta<\theta_0$}.
\end{array}
\right.
\end{equation}
% \end{itemize}

Maxima are then located using the algorithm described in \cite{Morhac1997Efficient}, 
and their position in the two-dimensional matrix (Fig. \ref{fig:spid}(c)) is used as starting point for all subsequently generated identification lines, making crucial the choice of $\theta_0$.

The operation of projection/localization is then repeated in order to cover the full
matrix, varying $\theta$ from $\theta_0$ to $90$\textdegree, and then from $\theta_0$ to $0$\textdegree~
by steps of $\delta \theta$. In practice, $\delta \theta$ and $\alpha$ slightly depend on $\theta$, and can
be modified by the user in order to adapt the algorithm to a specific situation (very low statistics for example).
Each new point $P(x_p,y_p)$ is associated to the line $Z$, so far containing $n_Z$ points and whose end point coordinates are
 $(x_Z,y_Z)$, if:
\begin{equation}
\left\{
\begin{array}{ll}
   | y_{p} - y_Z | < \delta y, &\text{ for $n_Z<10$},\\
   | y_{p} - f_Z(x_p) | < \delta y, &\text{ for $n_Z\geq 10$},
\end{array}
\right.
\end{equation}
with $f_Z(x) = a_Z^0\times(x+a_Z^1)^{-a_Z^2}$ a function fitted to the $n_Z$ points already associated to the line $Z$, 
and $\delta y = y_Z\times Z^{-1}$. 
%For Si-CsI(Tl), $f_Z(x) = a_Z^0\times(a_Z^1\times(x^{a_Z^2}+a_Z^3)^{-1}-1)$, while 
% in the other cases.
The choices of $f_Z(x)$ and $\delta y$ are purely phenomenological.

Once the spiderweb is woven (see Fig. \ref{fig:spid}(d)), identification lines that do not respect the following criteria:
\begin{equation}
\left\{
\begin{array}{l}
  n_Z>10\\
  a_Z^1<3000\\
  0.35<a_Z^2<1
\end{array}
\right.
\label{eq:crit}
\end{equation}
are rejected, where $n_Z$ is the final number of points associated to the line $Z$, and ($a_Z^0$, $ a_Z^1$, $a_Z^2$) 
are the parameters of $f_Z(x)$. This procedure aims at eliminating lines with a form completely incoherent
with the Bethe-Bloch formula, without being too restrictive in order for this method to be applicable to
different types of identification matrices. %The identification grid is then built from each individual functions $f_Z(x)$. 

It is then possible to build the identification grid from each of the
individual functions $f_Z(x)$, 
either limited to the range where peaks were found (see Fig.\ref{fig:coupl}(a) for example), 
or extrapolated over the whole residual energy range (see Fig.\ref{fig:sicsi}(a) for example).

\subsection{Implementation}
In order to run the algorithm presented above, several input parameters should be provided by the user:
the pedestal coordinates $(x_0,y_0)$, the first projection angle $\theta_0$, and the binning parameter $\beta$.
These parameters are generally difficult to estimate, which makes our method unusable without an implementation
in a suitable graphical environment.
It has therefore been included in the identification grid editor of KaliVeda \cite{kaliveda}, 
which is a graphical user interface dedicated to the creation and editing
of identification grids developed initially within the INDRA Collaboration \cite{Pouthas1995418}.

The pedestal coordinates $(x_0,y_0)$ can be set by the user with a simple click on the \Dee matrix.
Our algorithm needs also another point, $(X_0,Y_0)$, situated approximately 
on the middle of a high-$Z$ line; and the knowledge of the
charge $Z_0$ associated to this line. 
The values of $\theta_0$ and $\beta$ are then calculated as follow :
\begin{equation}
\tan(\theta_0) = \frac{Y_0-y_0}{X_0-x_0}
\end{equation}
\begin{equation}
\beta = \frac{1}{20 Z_0}\sqrt{(X_0-x_0)^2 + (Y_0-y_0)^2}.
\end{equation}

Thanks to the implementation in a ``user friendly'' graphical environment, our method needs only two mouse-clicks
from the user to calculate all initialization parameters, making it very easy to use.

\subsection{Examples of use}
The present method has been initially developed to treat INDRA Si-CsI(Tl) matrices \cite{Pouthas1995418}.
Since it uses as little information as possible on the exact form of $Z$ lines,
it can be applied
to different types of identification telescopes. Here are several examples of use on
Si-CsI(Tl), IC-Si, Si-Si matrices; and also matrices from Pulse Shape Analysis of the charge signal in silicon detectors.

\begin{figure*}[h]
\begin{center}
\includegraphics[width=.49\linewidth]{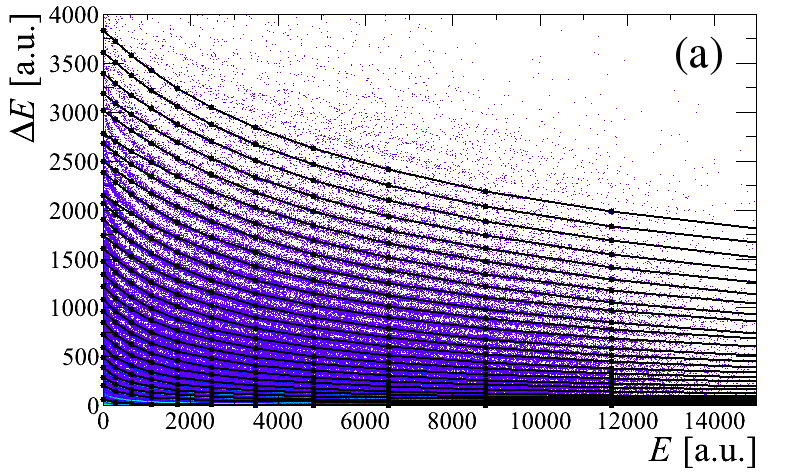}
\includegraphics[width=.49\linewidth]{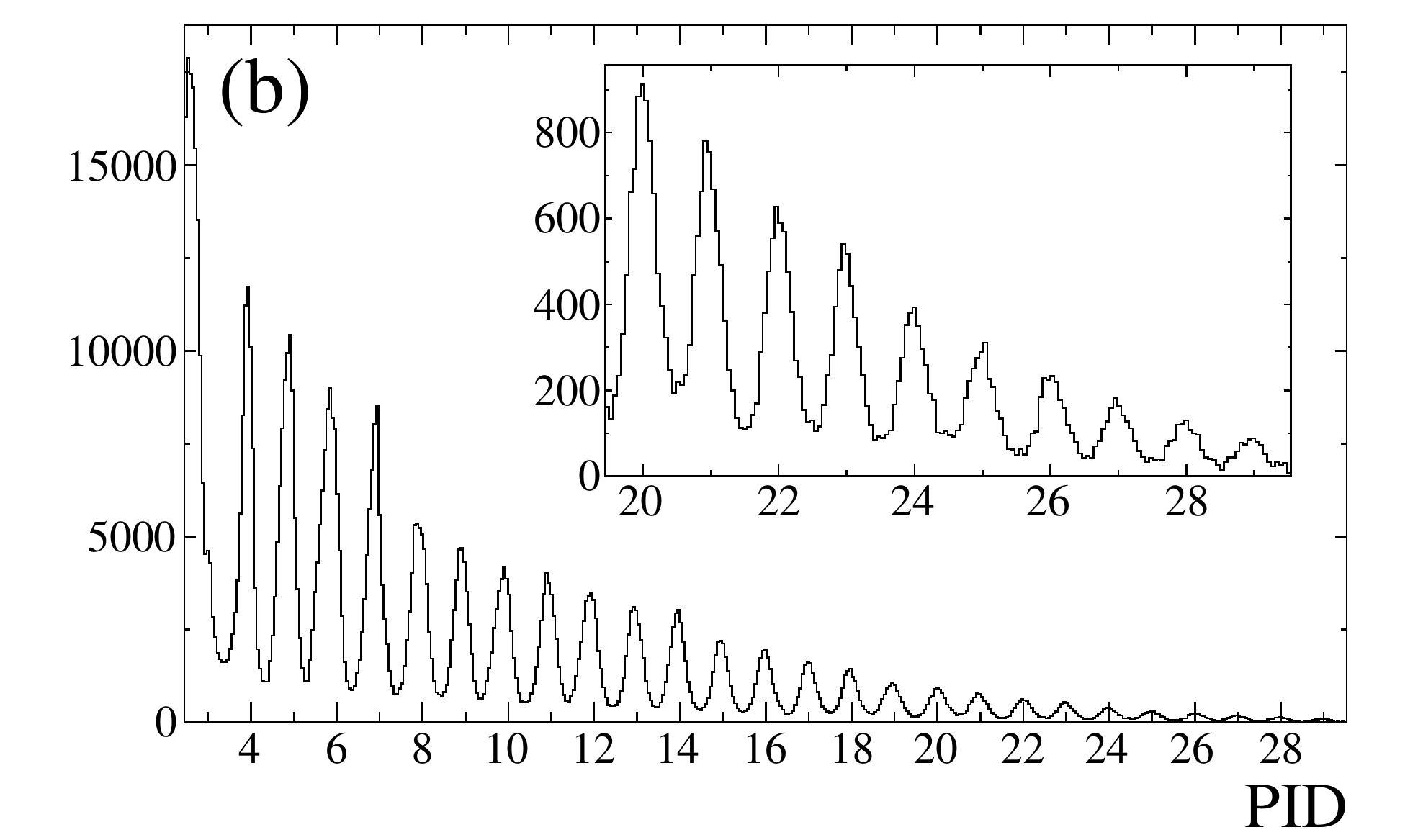}
\caption{Application of the SPIDER Identification method on a Si-CsI(Tl) matrix of INDRA ($\theta\sim17$\textdegree). PID stands for particle identification obtained after linearization of the matrix with the identification grid.
Data come from the Ta+Zn at 39\Mev reaction measured at GANIL.\label{fig:sicsi}}
\end{center}
\end{figure*}

Fig. \ref{fig:sicsi} presents the result of the SPIDER Identification on a \Dee matrix coming from
a Si-CsI(Tl) telescope of the INDRA multidetector. In this example, all lines
up to $Z=29$ were found by our algorithm (Fig. \ref{fig:sicsi}(a)). Low-$Z$ lines ($Z\lesssim15$)
are generally well reproduced over the whole residual energy range. For $Z\gtrsim15$ the
low energy part is systematically underestimated. In general, this part is not determined directly by the
localization algorithm but extrapolated using individual fitting functions. 
Nevertheless this grid provide a satisfying particle identification (Fig. \ref{fig:sicsi}(b)), but must be
slightly modified ``by hand'' before being used to make a definitive particle identification.
Extracted lines can also be used, without individual extrapolations, as input to constrain a fit using some functional.
This point is discussed in Section \ref{sec:fit}.

\begin{figure*}[h]
\begin{center}
\includegraphics[width=.49\linewidth]{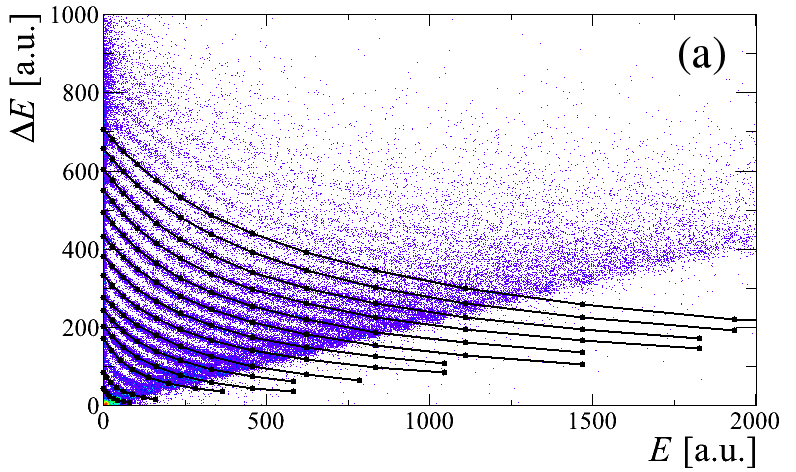}
\includegraphics[width=.49\linewidth]{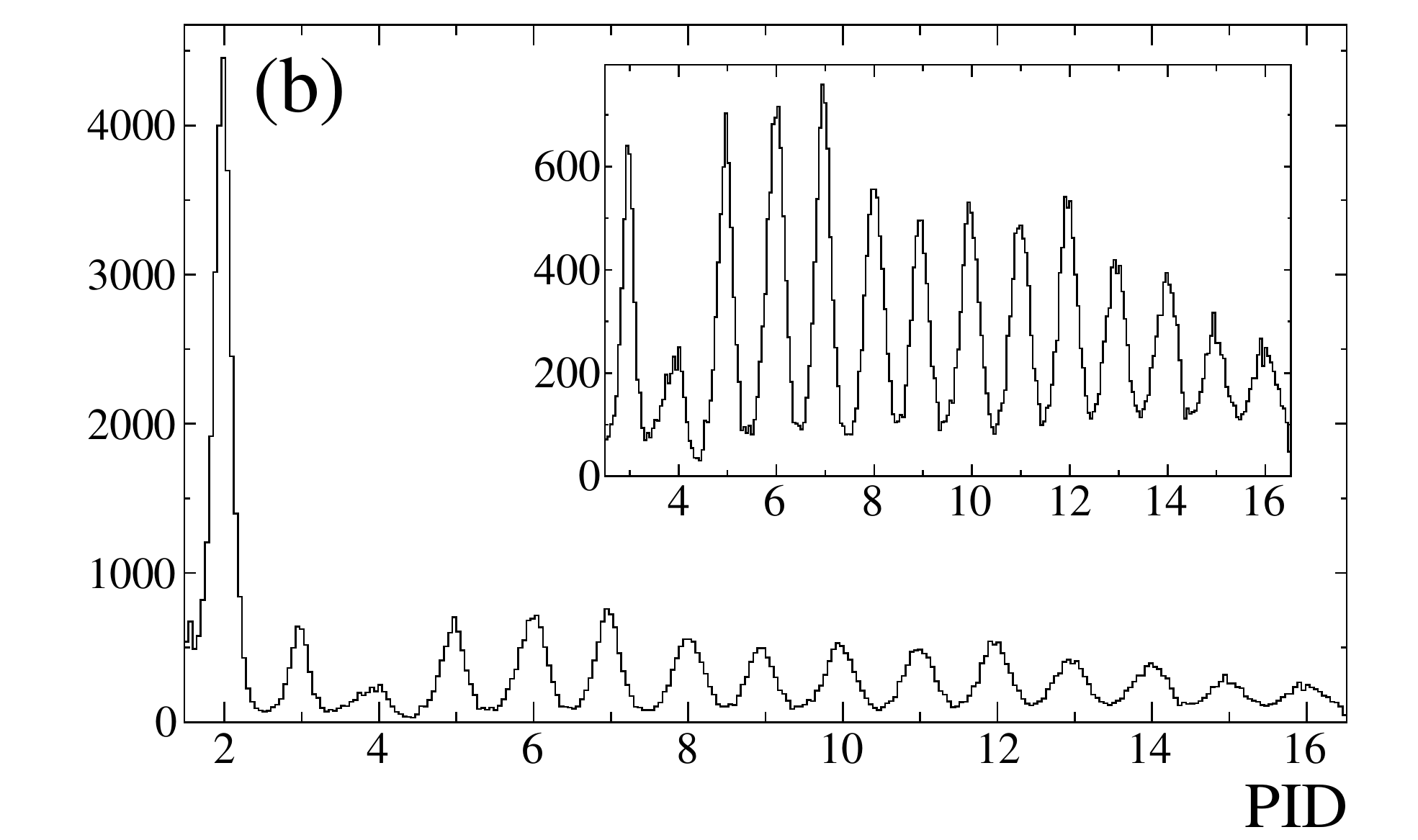}
\caption{Same as Fig. \ref{fig:sicsi} for IC-Si matrix of INDRA ($\theta\sim17$\textdegree). 
Data come from the Ta+Zn at 39\Mev reaction measured at GANIL.\label{fig:cisi}}
\end{center}
\end{figure*}

Fig. \ref{fig:cisi} presents the results of the SPIDER Identification on a \Dee matrix coming from
an INDRA IC-Si telescope.
In this kind of matrix, $Z$ lines are generally broad due to the poorer energy resolution of such large-area ionization chambers operated at low pressure, 
and rarely homogeneously populated. Nevertheless, our algorithm has extracted ridge lines from
$Z=2$ to $Z=16$ (Fig. \ref{fig:cisi}(a)), providing a good charge identification (Fig. \ref{fig:cisi}(b)).
It can be noted that, in this example, the $Z=4$ line was not generated because it did not satisfy the criteria of Eq. \eqref{eq:crit}.

\begin{figure*}[t]
\begin{center}
\includegraphics[width=.49\linewidth]{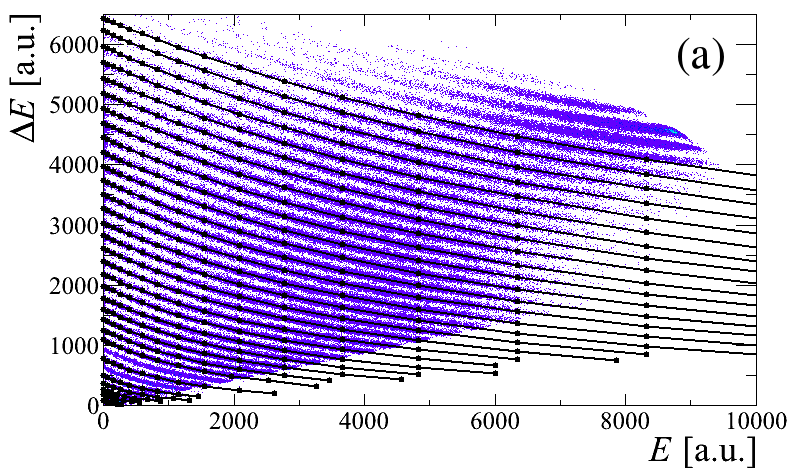}
\includegraphics[width=.49\linewidth]{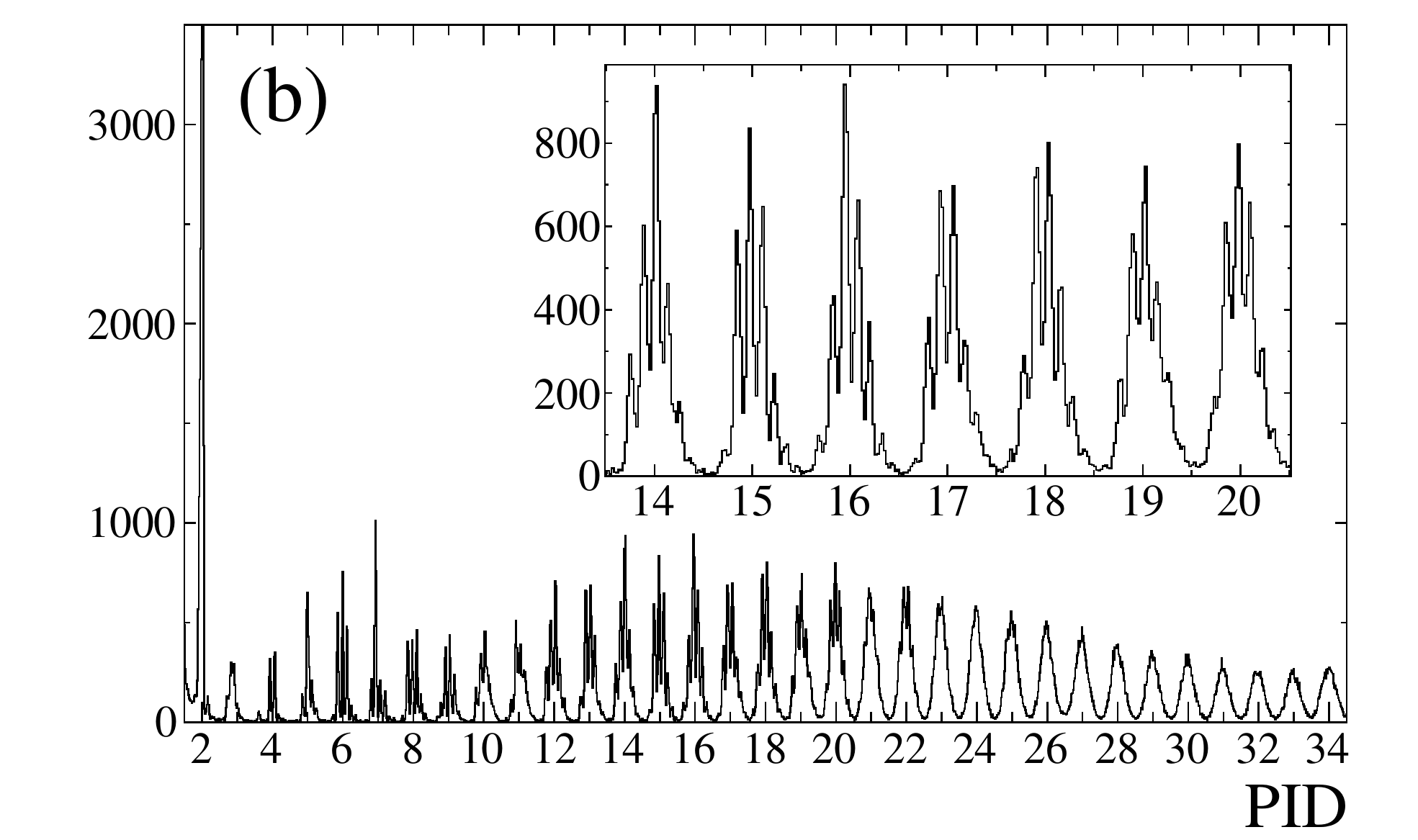}
\caption{Same as Fig. \ref{fig:sicsi} for Si-Si matrix of FAZIA ($\theta\sim7$\textdegree). 
Data come from the Kr+Sn at 35\Mev reaction measured at LNS \cite{PhysRevC.87.054607,PhysRevC.88.064607}.\label{fig:sisi}}
\end{center}
\end{figure*}

The result of the SPIDER Identification on a \Dee matrix coming from
a Si-Si telescope of FAZIA \cite{epjaFazia} is presented in Fig. \ref{fig:sisi}. 
In this kind of matrix, ridge lines corresponding to different elements are clearly separated thanks to
the very good quality of FAZIA silicon detectors. In addition, the line curvature is quite slight,
making the SPIDER Identification very efficient (Fig. \ref{fig:sisi}(a)). 
Even if a few lines ($Z=7$ and $Z=9$) have not been generated, the grid provides a very 
good charge identification, up to $Z=34$ (Fig. \ref{fig:sisi}(b)). For each integer value of $Z$, several peaks 
appear on the identification spectrum up to $Z\sim20$. These peaks correspond to different
isotopes of each element, and allow to discriminate particles of different masses.
The charge-identification grid can then be used as a starting point to generate a mass-identification
grid. This point is discussed in sec. \ref{sec:iso}.

\begin{figure*}[t]
\begin{center}
\includegraphics[width=.49\linewidth]{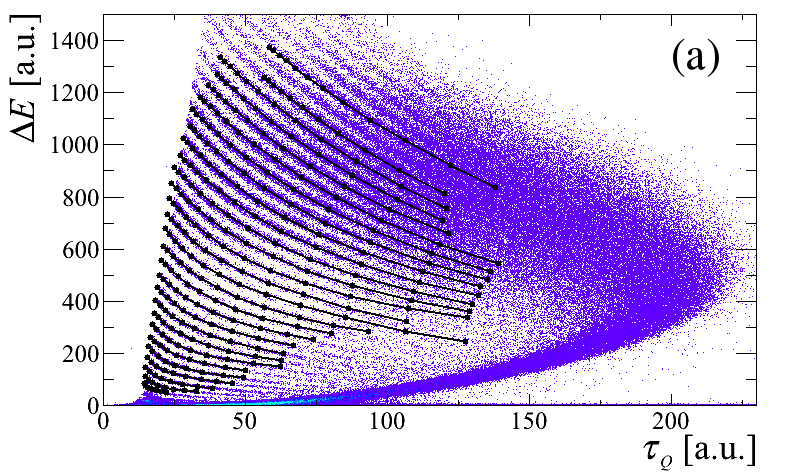}
\includegraphics[width=.49\linewidth]{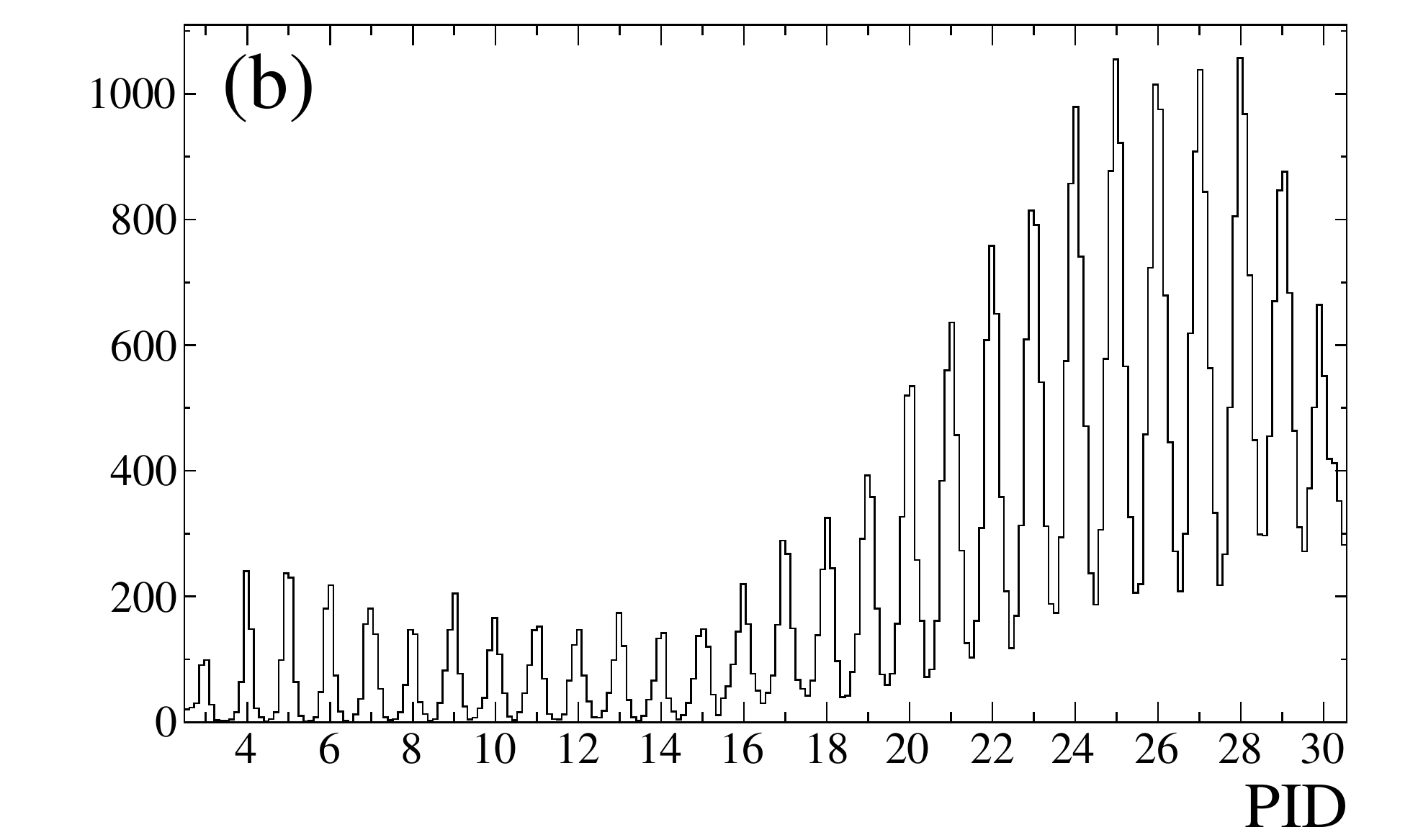}
\caption{Same as Fig. \ref{fig:sicsi} for energy versus rise time of the charge signal in the first silicon layer matrix of FAZIA ($\theta\sim7$\textdegree). 
Data come from the Kr+Sn at 35\Mev reaction measured at LNS \cite{PhysRevC.87.054607,PhysRevC.88.064607}.\label{fig:psa}}
\end{center}
\end{figure*}

Previous examples all concern the \Dee method. Identification matrices obtained by Pulse
Shape Analysis of the charge signal in FAZIA silicon detectors \cite{epjaFazia} present a form quite similar to that
obtained with the \Dee method. Since the SPIDER Identification method uses
little \textit{a priori} information on 
the ridge line form, it can also be applied to this type of matrix. 
In the example shown on Fig. \ref{fig:psa},
the generated grid provide a satisfying charge identification from $Z=3$ up to $Z=30$ (Fig. \ref{fig:psa}(b)), 
even if identification lines do not cover the full matrix range (Fig. \ref{fig:psa}(a)).

\subsection{Coupling with a fitting procedure}
\label{sec:fit}
In previous examples, $Z$-lines generated by the SPIDER identification, possibly extrapolated using individual functions,
are directly used to identify particles. This procedure is efficient but does not allow to extrapolate the identification to higher $Z$.
These lines, without extrapolation, can also be used as input to fit functional parameters.
In the present example, we used the functional proposed in \cite{TassanGot2002New}.

\begin{figure*}[t]
\begin{center}
\includegraphics[width=.49\linewidth]{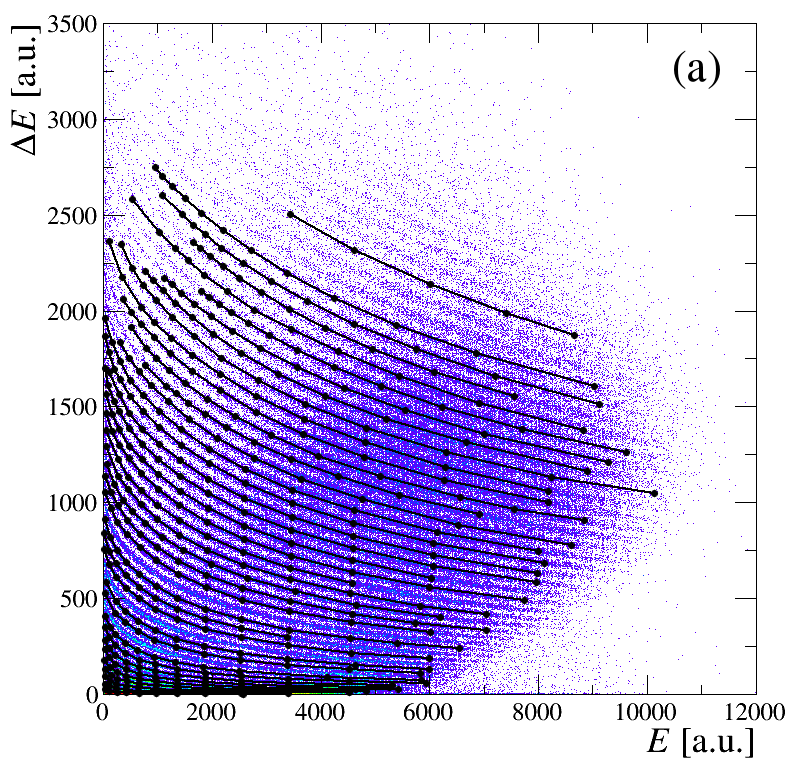}
\includegraphics[width=.49\linewidth]{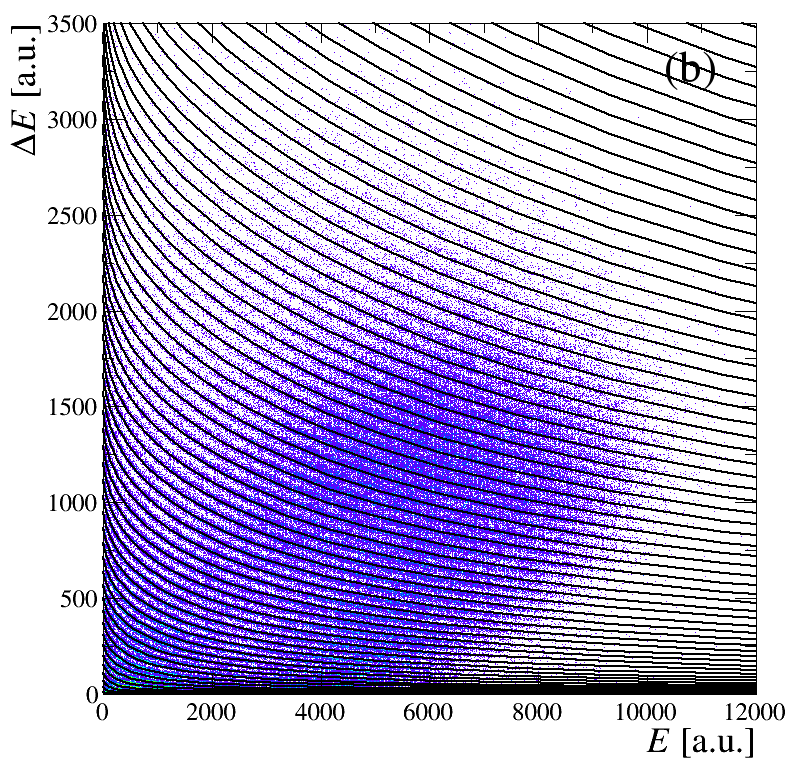}\\
\includegraphics[width=.96\linewidth]{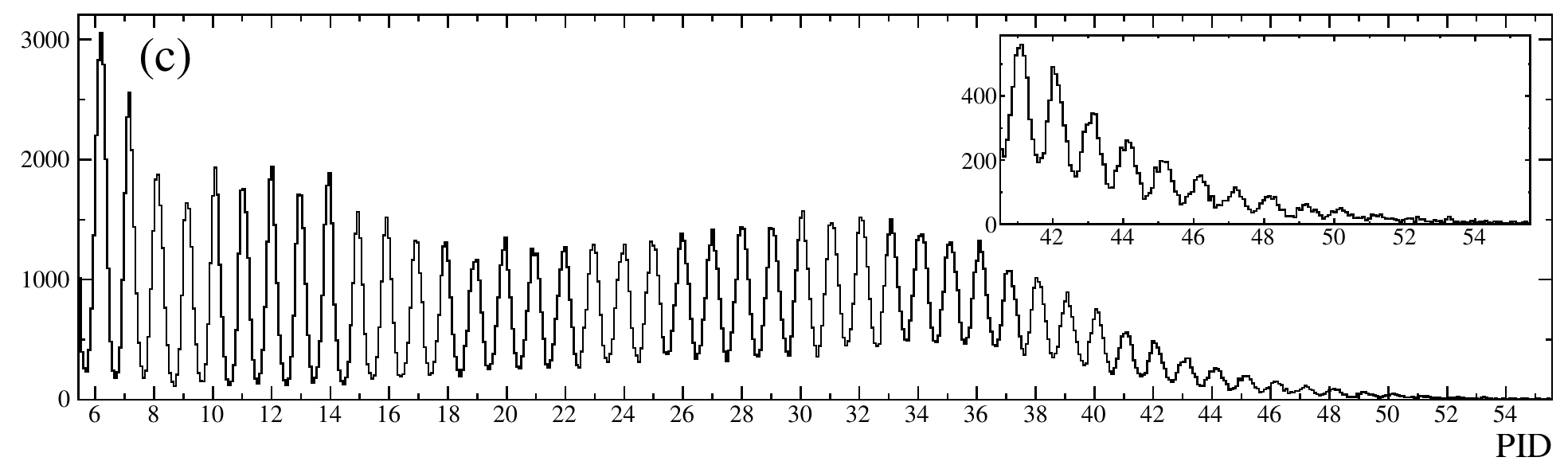}
\caption{
Example of coupling between the SPIDER Identification method and a functional fit \cite{TassanGot2002New} on a Si-CsI(Tl) matrix of INDRA ($\theta\sim8.5$\textdegree). 
Data come from the Ta+Zn at 39\Mev reaction measured at GANIL.
\label{fig:coupl}
}
\end{center}
\end{figure*}

Raw $Z$-lines obtained with the SPIDER method on an INDRA Si-CsI(Tl) matrix are presented in Fig. \ref{fig:coupl} (a).
These lines, which do not cover the whole residual energy range, are used to fit the 9 parameters of the
functional \cite{TassanGot2002New}. The result of the fit is shown on Fig. \ref{fig:coupl}(b). 
It can be seen that a satisfactory agreement is obtained for all charges and over the full matrix.
The quality of the charge identification can be checked on Fig. \ref{fig:coupl}(c):
a good charge identification is achieved up to $Z\sim50$, even if the statistics for high $Z$ is very poor.
The coupling between the SPIDER method and a functional fit allows to obtain a full charge identification
in a very short time. This procedure was used during the data reduction of the INDRA experiment presented in \cite{franklandIWMEC2014}.

\section{Isotopic identification}
\label{sec:iso}

The method presented above, and its possible coupling with a fitting procedure, facilitate a lot 
the extraction of $Z$-identification lines. As illustrated in Fig. \ref{fig:sisi}(b), the good quality of FAZIA
silicon detectors allows also isotopic identification of charged particles up to $Z\sim25$.
In order to carry out an isotopic identification, ridge lines corresponding to each $(Z,A)$ couple have
to be drawn. This is again a very fastidious task. We propose here a method to extract these
lines in a fully automatic way. The only input of the algorithm is a charge-identification grid that can be easily
generated using the SPIDER Identification method, which is particularly efficient for the case of 
Si-Si matrices (see Fig. \ref{fig:sisi}).

\begin{figure*}[t]
\begin{center}
      \includegraphics[width=0.74\linewidth]{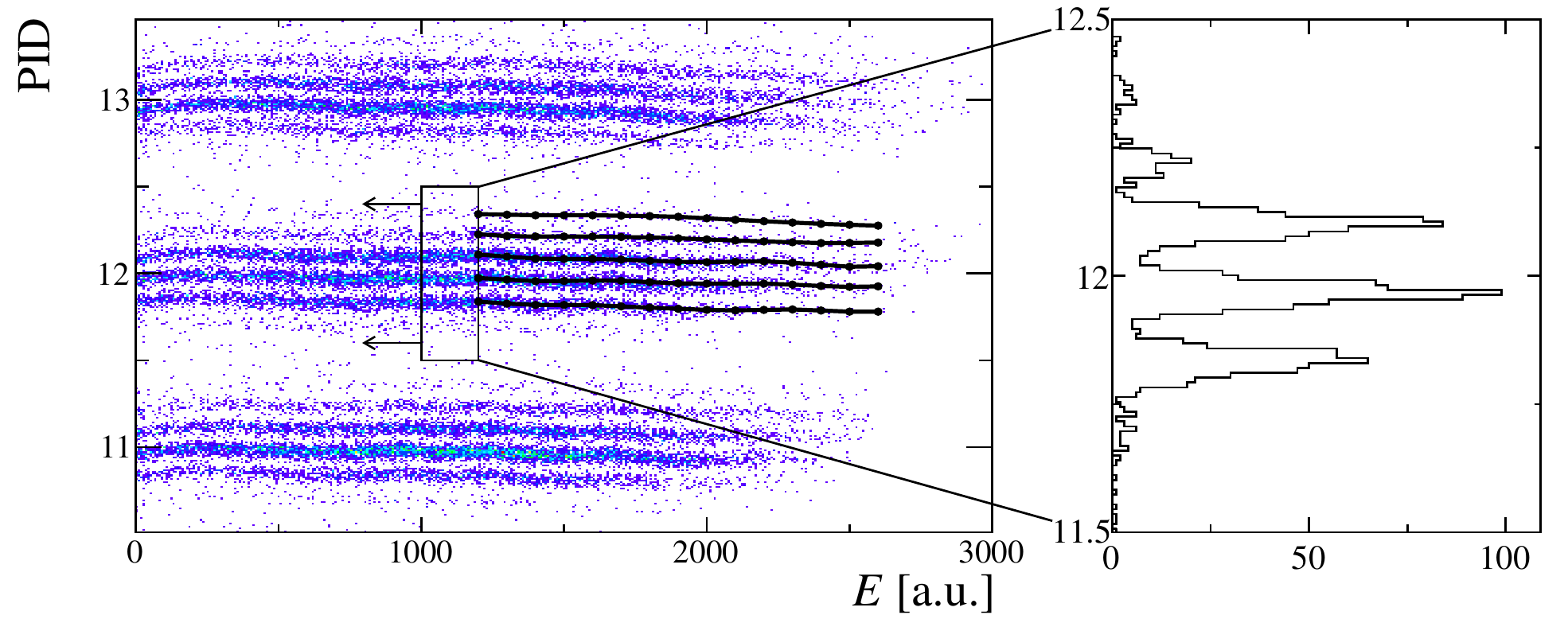}
\caption{Illustration of the isotopic identification method in PID-$E$ matrices.\label{fig:isomet}}
\end{center}
\end{figure*}

In \Dee matrices, lines corresponding to a given atomic number $Z$ are regularly spaced and populated. The scheme for isotopic lines is much more complex:
the relative population of each isotope depends strongly on the considered element, on the studied reaction, and
isotopes with short lifetimes (typically lower than $1$\,ns) are never detected.
The extraction of
isotopic $(Z,A)$-lines from the raw \Dee matrix is therefore very complex.
To simplify the treatment, we have to transform the matrix in
order to extract masses $Z$ by $Z$. The first stage of the algorithm is then to linearize the \Dee matrix according to
the $Z$-identification grid provided by the user, in order to obtain a PID$ -E$ matrix (Fig. \ref{fig:isomet}).
If the quality of the grid is good enough, $A$-lines associated to a given $Z$ are almost horizontal
and lie in the PID range $[Z$-$0.5$, $Z$+$0.5]$. 

\begin{figure*}[t]
\begin{center}
\includegraphics[width=.49\linewidth]{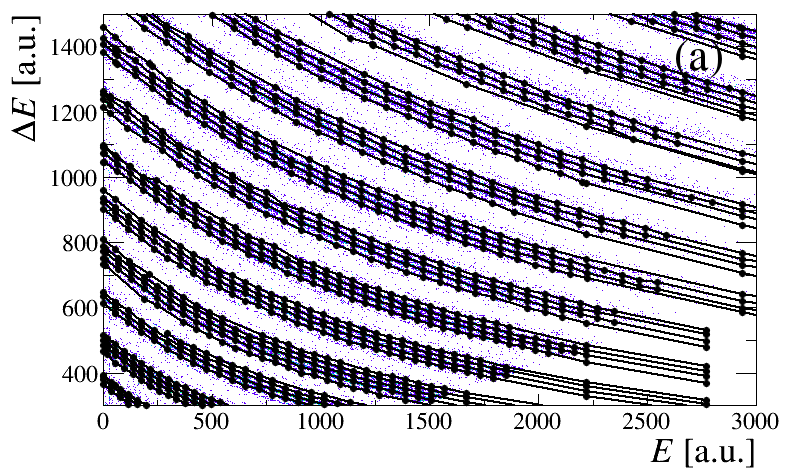}
\includegraphics[width=.49\linewidth]{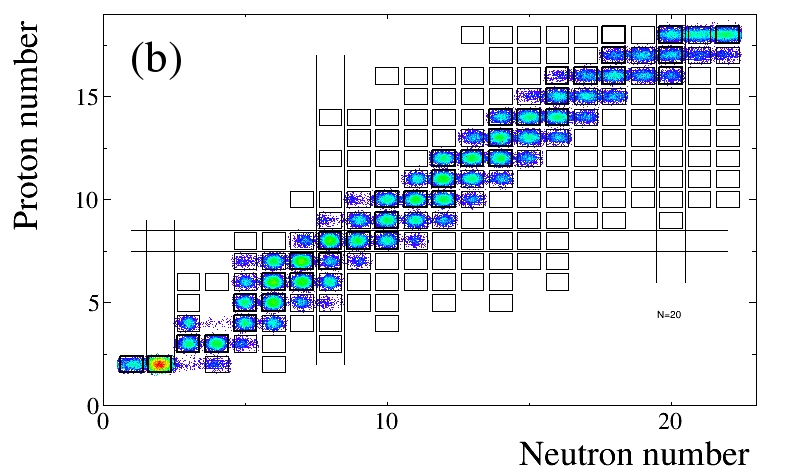}
%     \rfig{fig8a}{a}%
%     \lfig{fig8b}{b}%
\end{center}
\caption{
Result of the automatic isotopic line extraction applied to a Si-Si matrix of FAZIA ($\theta\sim7$\textdegree)
in the Ar+Sn at 25\Mev reaction: 
(a) isotopic identification grid; 
(b) isotopic distribution after linearization.}
\label{fig:isores}
\end{figure*}

All points contained in the range $\text{PID}=Z\pm0.5$ and $E=E_i\pm\delta E$
(rectangular box on Fig. \ref{fig:isomet}) are projected on the PID axis. Peaks corresponding to the intersection
of $A$-lines and the vertical line $E=E_i$ appear in the projection histogram. The binning of this projection
is set to $60$ bins whatever the $Z$ considered. Peaks are located and replaced in the PID$ -E$ matrix.
Each new point is simply associated to the closest $A$-line. The operation is then repeated by varying $E_i$
in order to cover the whole residual energy range, and for all $Z$ (Fig. \ref{fig:isomet}).
The obtained identification grid in the PID$ -E$  plane is finally transformed in the 
\Dee plane (Fig. \ref{fig:isores}(a)). The corresponding isotopic identification matrix is
presented in Fig. \ref{fig:isores}(b).

This algorithm has also been implemented in the identification grid editor of KaliVeda \cite{kaliveda}.

\section{Conclusion}
In this article, we proposed a new method (SPIDER identification) for generating $Z$ and $A$ identification grids in two-dimensional matrices. 
This method has been developed avoiding as much as possible the use of \textit{a priori} information on the exact form of identification lines,
in order to be applicable to a large variety of identification matrices. It has been successfully tested on various types of matrix obtained with \Dee and Pulse Shape Analysis techniques.
Particular attention has been paid to the implementation in a suitable graphical environment, so it needs only two mouse-clicks
from the user in order to calculate all initialization parameters.
Extracted lines can then be directly used to identify charged particles, 
set as an input of a functional fit, or used to extract isotopic lines in a fully automatic way.

\section{Acknowledgements}
The authors express their acknowledgement to the GANIL and LNS accelerator staff for their continued support during the experiments. We would especially like to thank the INDRA and FAZIA collaborations for allowing us to use their unpublished data. One of us (J.D.F.) would like to thank L. Tassan-Got for many fruitful discussions and for providing his code which is used in KaliVeda both for linearisation of the identification grids and for fitting to his identification functional(s). This work was supported by the Polish National Science Centre under Contract No. UMO-2014/14/M/ST2/00738 (COPIN-INFN Collaboration) and Contract No. UMO-2013/08/M/ST2/0025 (LEA-COPIGAL).

\bibliographystyle{unsrt}
\bibliography{paper}

\begin{thebibliography}{10}

\bibitem{WCI06}
Ph. Chomaz, F.~Gulminelli, W.~Trautmann, and S.~J. Yennello, editors.
\newblock {\em Dynamics and Thermodynamics with nuclear degrees of freedom},
  volume~30 of {\em Eur. Phys. J. A}.
\newblock Springer, 2006.

\bibitem{Borderie2008Nuclear}
B.~Borderie and M.~F. Rivet.
\newblock Nuclear multifragmentation and phase transition for hot nuclei.
\newblock {\em Prog. Part. Nucl. Phys.}, 61(2):551--601, October 2008.

\bibitem{STRACENER1990485}
D.W. Stracener, D.G. Sarantites, L.G. Sobotka, J.~Elson, J.T. Hood, Z.~Majka,
  V.~Abenante, A.~Chbihi, and D.C. Hensley.
\newblock Dwarf ball and dwarf wall: Design, instrumentation, and response
  characteristics of a 4$\pi$ csi(tl) plastic phoswich multidetector system for
  light charged particle and intermediate mass fragment spectrometry.
\newblock {\em Nuclear Instruments and Methods in Physics Research Section A:
  Accelerators, Spectrometers, Detectors and Associated Equipment}, 294(3):485
  -- 503, 1990.

\bibitem{Pouthas1995418}
J.~Pouthas, B.~Borderie, R.~Dayras, E.~Plagnol, M.F. Rivet, F.~Saint-Laurent,
  J.C. Steckmeyer, G.~Auger, C.O. Bacri, S.~Barbey, A.~Barbier, A.~Benkirane,
  J.~Benlliure, B.~Berthier, E.~Bougamont, P.~Bourgault, P.~Box, R.~Bzyl,
  B.~Cahan, Y.~Cassagnou, D.~Charlet, J.L. Charvet, A.~Chbihi, T.~Clerc,
  N.~Copinet, D.~Cussol, M.~Engrand, J.M. Gautier, Y.~Huguet, O.~Jouniaux, J.L.
  Laville, P.~Le Botlan, A.~Leconte, R.~Legrain, P.~Lelong, M.~Le Guay,
  L.~Martina, C.~Mazur, P.~Mosrin, L.~Olivier, J.P. Passerieux, S.~Pierre,
  B.~Piquet, E.~Plaige, E.C. Pollacco, B.~Raine, A.~Richard, J.~Ropert,
  C.~Spitaels, L.~Stab, D.~Sznajderman, L.~Tassan-got, J.~Tillier, M.~Tripon,
  P.~Vallerand, C.~Volant, P.~Volkov, J.P. Wieleczko, and G.~Wittwer.
\newblock {INDRA}, a 4$\pi$ charged product detection array at {GANIL}.
\newblock {\em Nucl. Instrum. Methods Phys. Res., Sect. A}, 357:418 -- 442,
  1995.

\bibitem{KWIATKOWSKI1995571}
K~Kwiatkowski, D.S Bracken, K.B Morley, J~Brzychczyk, E~Renshaw Foxford,
  K~Komisarcik, V.E Viola, N.R Yoder, J~Dorsett, J~Poehlman, N~Madden, and
  J~Ottarson.
\newblock The indiana silicon sphere 4$\pi$ charged-particle detector array.
\newblock {\em Nuclear Instruments and Methods in Physics Research Section A:
  Accelerators, Spectrometers, Detectors and Associated Equipment}, 360(3):571
  -- 583, 1995.

\bibitem{SARANTITES1996418}
D.G. Sarantites, P.-F. Hua, M.~Devlin, L.G. Sobotka, J.~Elson, J.T. Hood, D.R.
  LaFosse, J.E. Sarantites, and M.R. Maier.
\newblock The microball design, instrumentation and response characteristics of
  a 4$\pi$-multidetector exit channel-selection device for spectroscopic and
  reaction mechanism studies with gammasphere.
\newblock {\em Nuclear Instruments and Methods in Physics Research Section A:
  Accelerators, Spectrometers, Detectors and Associated Equipment}, 381(2):418
  -- 432, 1996.

\bibitem{PAGANO2004504}
A.~Pagano, M.~Alderighi, F.~Amorini, A.~Anzalone, L.~Arena, L.~Auditore,
  V.~Baran, M.~Bartolucci, I.~Berceanu, J.~Blicharska, J.~Brzychczyk,
  A.~Bonasera, B.~Borderie, R.~Bougault, M.~Bruno, G.~Cardella, S.~Cavallaro,
  M.B. Chatterjee, A.~Chbihi, J.~Cibor, M.~Colonna, M.~D'Agostino, R.~Dayras,
  E.~De Filippo, M.~Di Toro, W.~Gawlikowicz, E.~Geraci, F.~Giustolisi,
  A.~Grzeszczuk, P.~Guazzoni, D.~Guinet, M.~Iacono-Manno, S.~Kowalski, E.~La
  Guidara, G.~Lanzano, G.~Lanzalone, N.~Le Neindre, S.~Li, S.~Lo Nigro,
  C.~Maiolino, Z.~Majka, G.~Manfredi, T.~Paduszynski, M.~Papa, M.~Petrovici,
  E.~Piasecki, S.~Pirrone, R.~Planeta, G.~Politi, A.~Pop, F.~Porto, M.F. Rivet,
  E.~Rosato, F.~Rizzo, S.~Russo, P.~Russotto, M.~Sassi, G.~Sechi, V.~Simion,
  K.~Siwek-Wilczynska, I.~Skwira, M.L. Sperduto, J.C. Steckmeyer, L.~Swiderski,
  A.~Trifiro`, M.~Trimarchi, G.~Vannini, M.~Vigilante, J.P. Wieleczko,
  J.~Wilczynski, H.~Wu, Z.~Xiao, L.~Zetta, and W.~Zipper.
\newblock Fragmentation studies with the {CHIMERA} detector at {LNS} in
  {Catania}: recent progress.
\newblock {\em Nuclear Physics A}, 734:504 -- 511, 2004.

\bibitem{Wuenschel2009578}
S.~Wuenschel, K.~Hagel, R.~Wada, J.B. Natowitz, S.J. Yennello, Z.~Kohley,
  C.~Bottosso, L.W. May, W.B. Smith, D.V. Shetty, B.C. Stein, S.N. Soisson, and
  G.~Prete.
\newblock Nimrod-isis, a versatile tool for studying the isotopic degree of
  freedom in heavy ion collisions.
\newblock {\em Nuclear Instruments and Methods in Physics Research Section A:
  Accelerators, Spectrometers, Detectors and Associated Equipment}, 604(3):578
  -- 583, 2009.

\bibitem{epjaFazia}
R.~Bougault, G.~Poggi, S.~Barlini, B.~Borderie, G.~Casini, A.~Chbihi,
  N.~Le~Neindre, M.~P\^arlog, G.~Pasquali, S.~Piantelli, Z.~Sosin, G.~Ademard,
  R.~Alba, A.~Anastasio, S.~Barbey, L.~Bardelli, M.~Bini, A.~Boiano,
  M.~Boisjoli, E.~Bonnet, R.~Borcea, B.~Bougard, G.~Brulin, M.~Bruno,
  S.~Carboni, C.~Cassese, F.~Cassese, M.~Cinausero, L.~Ciolacu, I.~Cruceru,
  M.~Cruceru, B.~D'Aquino, B.~De~Fazio, M.~Degerlier, P.~Desrues, P.~Di~Meo,
  J.A. Due\~nas, P.~Edelbruck, S.~Energico, M.~Falorsi, J.D. Frankland,
  E.~Galichet, K.~Gasior, F.~Gramegna, R.~Giordano, D.~Gruyer, A.~Grzeszczuk,
  M.~Guerzoni, H.~Hamrita, C.~Huss, M.~Kajetanowicz, K.~Korcyl, A.~Kordyasz,
  T.~Kozik, P.~Kulig, L.~Lavergne, E.~Legouée, O.~Lopez, J.~\u0141ukasik,
  C.~Maiolino, T.~Marchi, P.~Marini, I.~Martel, V.~Masone, A.~Meoli, Y.~Merrer,
  L.~Morelli, F.~Negoita, A.~Olmi, A.~Ordine, G.~Paduano, C.~Pain, M.~Palka,
  G.~Passeggio, G.~Pastore, P.~Pawlowski, M.~Petcu, H.~Petrascu, E.~Piasecki,
  G.~Pontoriere, E.~Rauly, M.F. Rivet, R.~Rocco, E.~Rosato, L.~Roscilli,
  E.~Scarlini, F.~Salomon, D.~Santonocito, V.~Seredov, S.~Serra, D.~Sierpowski,
  G.~Spadaccini, C.~Spitaels, A.A. Stefanini, G.~Tobia, G.~Tortone, T.~Twaróg,
  S.~Valdré, A.~Vanzanella, E.~Vanzanella, E.~Vient, M.~Vigilante, G.~Vitiello,
  E.~Wanlin, A.~Wieloch, and W.~Zipper.
\newblock The {FAZIA} project in europe: {R\&D} phase.
\newblock {\em The European Physical Journal A}, 50(2), 2014.

\bibitem{Barlini2009644}
S.~Barlini, R.~Bougault, Ph. Laborie, O.~Lopez, D.~Mercier, M.~Parlog,
  B.~Tamain, E.~Vient, E.~Chevallier, A.~Chbihi, B.~Jacquot, and V.L. Kravchuk.
\newblock New digital techniques applied to a and z identification using pulse
  shape discrimination of silicon detector current signals.
\newblock {\em Nuclear Instruments and Methods in Physics Research Section A:
  Accelerators, Spectrometers, Detectors and Associated Equipment}, 600(3):644
  -- 650, 2009.

\bibitem{Bardelli2011272}
L.~Bardelli, M.~Bini, G.~Casini, P.~Edelbruck, G.~Pasquali, G.~Poggi,
  S.~Barlini, R.~Berjillos, B.~Borderie, R.~Bougault, M.~Bruno, S.~Carboni,
  A.~Chbihi, M.~D'Agostino, J.A.~Due\ nas, J.M. Gautier, F.~Gramegna, C.~Huss,
  A.J. Kordyasz, T.~Kozik, V.L. Kravchuk, N.~Le Neindre, O.~Lopez, I.~Martel,
  L.~Morelli, A.~Ordine, M.F. Rivet, E.~Rosato, E.~Scarlini, G.~Spadaccini,
  G.~Tobia, M.~Vigilante, and E.~Wanlin.
\newblock Progresses in the pulse shape identification with silicon detectors
  within the {FAZIA} collaboration.
\newblock {\em Nuclear Instruments and Methods in Physics Research Section A:
  Accelerators, Spectrometers, Detectors and Associated Equipment}, 654(1):272
  -- 278, 2011.

\bibitem{Goulding1964New}
F.~S. Goulding, D.~A. Landis, J.~Cerny, and R.~H. Pehl.
\newblock {A new particle identifier technique for Z = 1 and Z = 2 particles in
  the energy range $>$ 10 MeV}.
\newblock {\em Nuclear Instruments and Methods}, 31(1):1--12, December 1964.

\bibitem{Butler1970Identification}
G.~W. Butler, A.~M. Poskanzer, and D.~A. Landis.
\newblock {Identification of nuclear fragments by a combined time-of-flight,
  $\Delta$E-E technique}.
\newblock {\em Nuclear Instruments and Methods}, 89:189--198, December 1970.

\bibitem{TassanGot2002New}
L.~Tassan-Got.
\newblock {A new functional for charge and mass identification in $\Delta$E$-$E
  telescopes}.
\newblock {\em Nuclear Instruments and Methods in Physics Research Section B:
  Beam Interactions with Materials and Atoms}, 194(4):503--512, October 2002.

\bibitem{Benkirane1995Contextual}
A.~Benkirane, G.~Auger, D.~Bloyet, A.~Chbihi, and E.~Plagnol.
\newblock {A contextual image segmentation system using a priori information
  for automatic data classification in nuclear physics}.
\newblock {\em Nuclear Instruments and Methods in Physics Research Section A:
  Accelerators, Spectrometers, Detectors and Associated Equipment},
  355(2-3):559--574, February 1995.

\bibitem{MorhacIdentification}
Miroslav Morh\'a\^c and Martin Veselský.
\newblock Identification of isotope lines in two-dimensional spectra of nuclear
  multifragmentation.
\newblock {\em Nuclear Instruments and Methods in Physics Research Section A:
  Accelerators, Spectrometers, Detectors and Associated Equipment}, 592(3):434
  -- 450, 2008.

\bibitem{Morelli2010Automatic}
L.~Morelli, M.~Bruno, G.~Baiocco, L.~Bardelli, S.~Barlini, M.~Bini, G.~Casini,
  M.~D'Agostino, M.~Degerlier, F.~Gramegna, V.~L. Kravchuk, T.~Marchi,
  G.~Pasquali, and G.~Poggi.
\newblock {Automatic procedure for mass and charge identification of light
  isotopes detected in CsI(Tl) of the GARFIELD apparatus}.
\newblock {\em Nuclear Instruments and Methods in Physics Research Section A:
  Accelerators, Spectrometers, Detectors and Associated Equipment},
  620(2-3):305--313, August 2010.

\bibitem{Dudouet2013Comparison}
J.~Dudouet, D.~Juliani, M.~Labalme, J.~C. Ang\'{e}lique, B.~Braunn, J.~Colin,
  D.~Cussol, Ch~Finck, J.~M. Fontbonne, H.~Gu\'{e}rin, P.~Henriquet,
  J.~Krimmer, M.~Rousseau, and M.~G. Saint-Laurent.
\newblock {Comparison of two analysis methods for nuclear reaction measurements
  of 12C +12C interactions at 95MeV/u for hadron therapy}.
\newblock {\em Nuclear Instruments and Methods in Physics Research Section A:
  Accelerators, Spectrometers, Detectors and Associated Equipment},
  715:98--104, July 2013.

\bibitem{Brun1997ROOT}
R.~Brun.
\newblock {ROOT} — an object oriented data analysis framework.
\newblock {\em Nuclear Instruments and Methods in Physics Research Section A:
  Accelerators, Spectrometers, Detectors and Associated Equipment},
  389(1-2):81--86, April 1997.

\bibitem{kaliveda}
KaliVeda data~analysis toolkit.
\newblock http://indra.in2p3.fr/KaliVedaDoc.

\bibitem{Morhac1997Efficient}
Miroslav Morh\'{a}\v{c}, J\'{a}n Kliman, Vladislav Matou\v{s}ek, Martin
  Veselsk\'{y}, and Ivan Turzo.
\newblock {Efficient one- and two-dimensional gold deconvolution and its
  application to $\gamma$-ray spectra decomposition}.
\newblock {\em Nuclear Instruments and Methods in Physics Research Section A:
  Accelerators, Spectrometers, Detectors and Associated Equipment},
  401(2-3):385--408, 1997.

\bibitem{PhysRevC.87.054607}
S.~Barlini, S.~Piantelli, G.~Casini, P.~R. Maurenzig, A.~Olmi, M.~Bini,
  S.~Carboni, G.~Pasquali, G.~Poggi, A.~A. Stefanini, R.~Bougault, E.~Bonnet,
  B.~Borderie, A.~Chbihi, J.~D. Frankland, D.~Gruyer, O.~Lopez, N.~Le~Neindre,
  M.~P\^arlog, M.~F. Rivet, E.~Vient, E.~Rosato, G.~Spadaccini, M.~Vigilante,
  M.~Bruno, T.~Marchi, L.~Morelli, M.~Cinausero, M.~Degerlier, F.~Gramegna,
  T.~Kozik, T.~Twar\'og, R.~Alba, C.~Maiolino, and D.~Santonocito.
\newblock Isospin transport in ${}^{84}$kr+${}^{112,124}$sn collisions at fermi
  energies.
\newblock {\em Phys. Rev. C}, 87:054607, May 2013.

\bibitem{PhysRevC.88.064607}
S.~Piantelli, G.~Casini, P.~R. Maurenzig, A.~Olmi, S.~Barlini, M.~Bini,
  S.~Carboni, G.~Pasquali, G.~Poggi, A.~A. Stefanini, S.~Valdr\`e, R.~Bougault,
  E.~Bonnet, B.~Borderie, A.~Chbihi, J.~D. Frankland, D.~Gruyer, O.~Lopez,
  N.~Le~Neindre, M.~P\^arlog, M.~F. Rivet, E.~Vient, E.~Rosato, G.~Spadaccini,
  M.~Vigilante, M.~Bruno, T.~Marchi, L.~Morelli, M.~Cinausero, M.~Degerlier,
  F.~Gramegna, T.~Kozik, T.~Twar\'og, R.~Alba, C.~Maiolino, and D.~Santonocito.
\newblock $n$ and $z$ odd-even staggering in kr+sn collisions at fermi
  energies.
\newblock {\em Phys. Rev. C}, 88:064607, Dec 2013.

\bibitem{franklandIWMEC2014}
J.D. Frankland, D.~Gruyer, E.~Bonnet, and A.~Chbihi.
\newblock Comparison of radial flow effects on partitions of multifragmenting
  sources formed in symmetric and asymmetric central collisions.
\newblock {\em EPJ Web of Conferences}, 88:00009, 2015.

\end{thebibliography}

\end{document}